\documentclass{ws-p8-50x6-00}
\usepackage{amsmath}
\usepackage{amssymb}
\usepackage{physics}
\usepackage{pennames}

\begin{document}

\title{MOMENTS OF THE CHARGED-PARTICLE MULTIPLICITY DISTRIBUTION IN Z DECAYS AT LEP}

\author{D. J. MANGEOL}

\address{HEFIN, University of Nijmegen/NIKHEF,\\ P.O. Box 9010, 6500 GL Nijmegen, 
The Netherlands\\E-mail: Dominique.Mangeol@cern.ch} 

\address{\em on behalf of the L3 collaboration}

\maketitle\abstracts{The charged-particle multiplicity distribution and its 
moments have been measured, for all hadronic as well as for light-quark and b-quark events 
in $\Pep\Pem$ collisions at the Z mass. The $H_q$ moments derived from the 
charged-particle multiplicity distribution are known to exhibit quasi-oscillations 
when plotted versus the order of the moment. 
This behavior is predicted by the NNLLA of perturbative QCD for the parton level and, 
under the assumption of LPHD, also for the hadron level. Using the jet 
multiplicity distributions in order to 
vary the dependence on the LPHD hypothesis, we find, however, that at our energy the 
oscillations only appear for non-perturbative scales. In the absence of confirmation 
of pQCD, we investigate a more phenomenological answer in the possibility that the 
features seen in the $H_q$ behavior could be due to the fact that the charged-particle 
multiplicity derives from a superposition of final states related to the topology of 
the events. Therefore, the analysis is repeated using charged-particle multiplicity 
distributions originating from 2-jet and 3-jet events for the full, light- and 
b-quark samples.}

\section{Introduction}

Although the number of charged particles is only a global measure of the 
characteristics of the final state of a high-energy collision, it has proved 
a fundamental tool in the study of particle production. Independent emission 
of single particles leads to a Poissonian multiplicity distribution. 
Deviations from this shape, therefore, reveal correlations~\cite{kittel}.
The shape of the charged-particle multiplicity distribution analysed 
with the ratio of cumulant factorial moments to 
factorial moments\cite{dremin1}, $H_q$, is known to reveal quasi-oscillations~\cite{SLD}, 
when plotted versus the order q, with a first minimum at $q=5$.

The usual way to interpret this result is to refer to perturbative 
QCD, which provides us with calculations for the $H_q$ of the parton multiplicity 
distribution~\cite{dremin2}. 
The Next to Next to Leading Logarithm
Approximation (NNLLA), which has the most accurate treatement of energy-momentum
conservation, predicts for the $H_q$ a negative first minimum near 5 followed by 
quasi-oscillations.
This behavior may be expected for the charged-particle multiplicity distribution
under the Local Parton-Hadron Duality (LPHD) hypothesis, which assumes that the 
hadronization does not distort the shape of the  multiplicity distribution.

However, this result can also be interpreted in a more phenomenological way by viewing  
the shape of the charged-particle multiplicity distribution as a superposition of 
different types of event like 2-jet and 3-jet events\cite{shoulder}.
This can be investigated using rather simple parametrizations, as a weighted 
sum of two Negative Binomial Distributions (NBD), each NBD carrying parameters 
(mean, $\bar{n}$ and dispersion, $D$) taken from the experimental 2-jet or 3-jet 
charged-particle multiplicity distributions and using as relative weight the 2-jet fraction 
($\alpha_{\mathrm{2jet}}$),
$\mathrm{2NBD}_{\mathrm{full sample}}=\alpha_{\mathrm{2jet}}\mathrm{NBD}_{\mathrm{2jet}}+
                                (1-\alpha_{\mathrm{2jet}})\mathrm{NBD}_{\mathrm{3jet}}$.
A similar parametrization can also be tested using light- and b-quark events\cite{flavour}, 
instead of 2-jet and 3-jet events.

The test of the two approaches is done by  measuring charged-particle multiplicity 
distributions and their moments for the full, light- and b-quark samples. 
These samples were also subdivided into 2-jet and 
3-jet events obtained from various $y_{\mathrm{cut}}$ values.

This analysis is based on data collected by the L3 detector\cite{det1} in 1994 and 1995 
at the energy of the $\mathrm{Z}$. The data sample corresponds to approximately 
one million selected hadronic events. A b-tag algorithm is used to discriminate
between light-(udsc) and b-quark events\cite{btag}.
Furthermore, the resulting multiplicity
distributions are fully corrected for selection, detector 
inefficiencies\cite{tune,sil3,susinno} and light- or b-quark purity.
It has been shown that the $H_q$ moments are very sensitive to truncation\cite{trunc}. 
Since we want to compare a large variety of multiplicity distributions, 
we have to make sure that all distributions are affected by the truncation in 
the same way. Therefore, the truncation is defined as the fraction
of events removed in the tail of the charged-particle multiplicity distribution of
the full sample (events with multiplicity larger than 48 are removed), and the 
same fraction of events is then removed in all charged-particle multiplicity 
distributions studied.

\section{Test of the pQCD approach}   
\begin{figure}[htbp]
  \begin{center}
    \includegraphics[width=5.8cm,height=6cm]{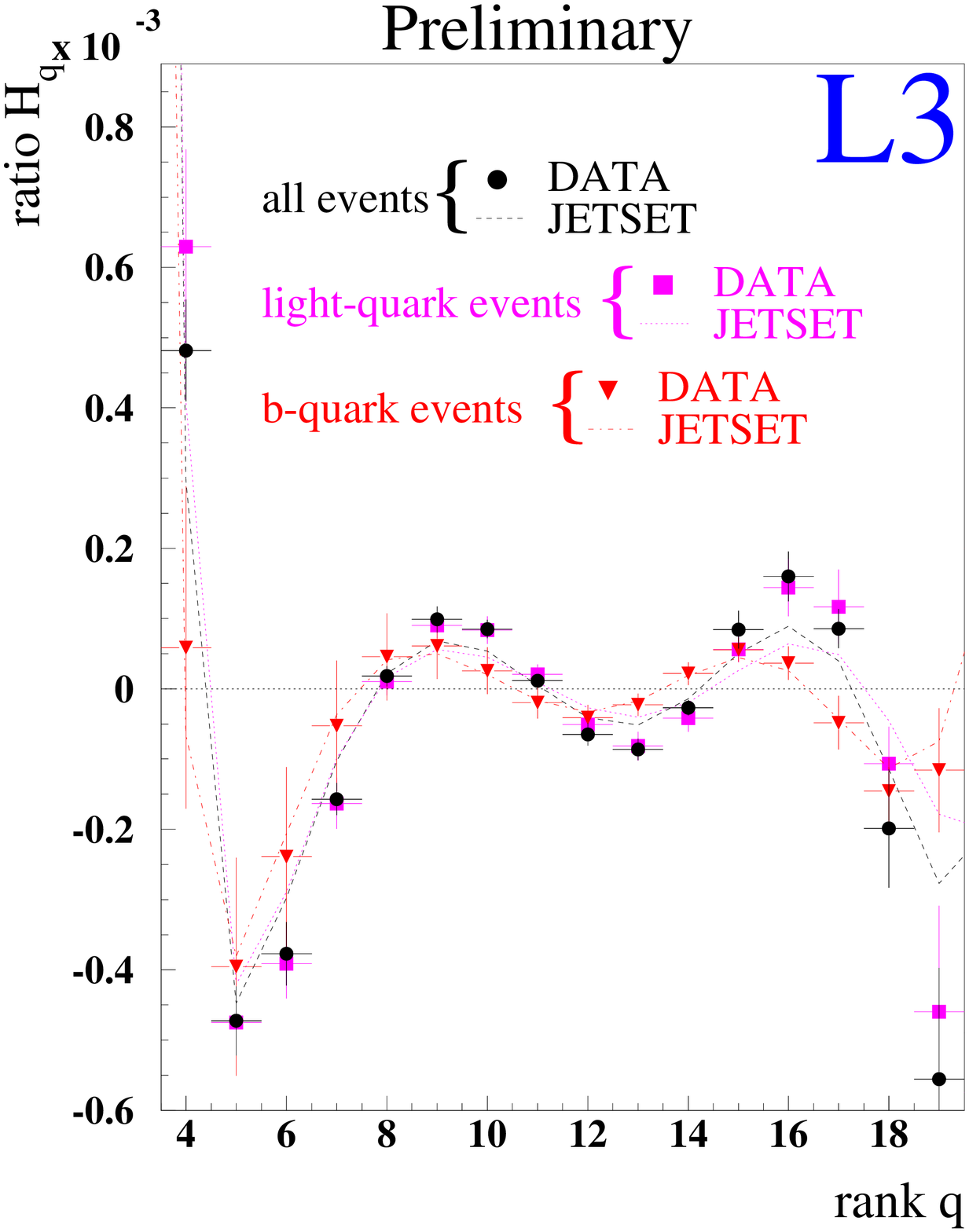}
    \includegraphics[width=5.8cm,height=6cm]{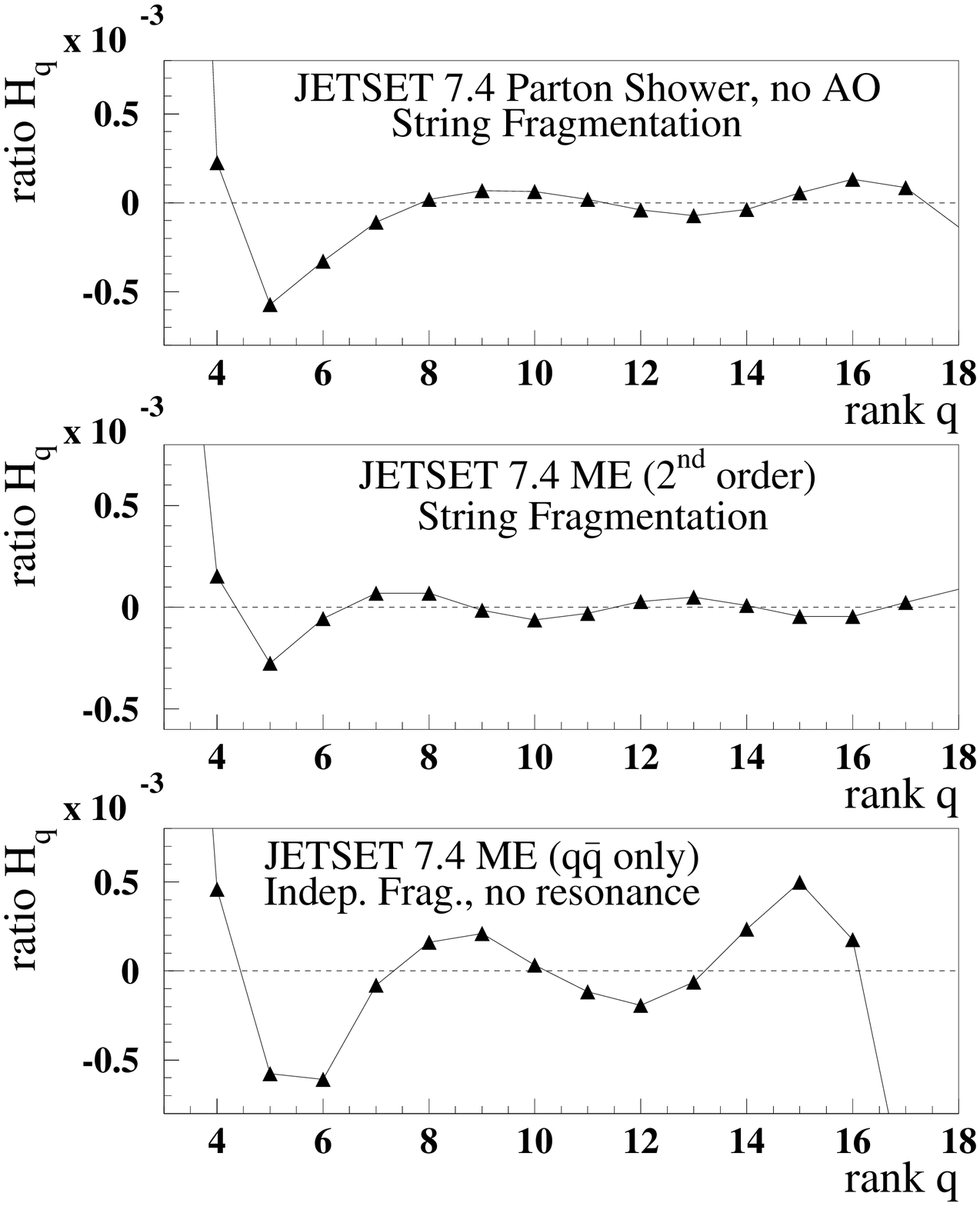}
\begin{minipage}[htbp]{5.4cm}
 \caption{$H_q$ of the charged particle multiplicity distribution.}
 \label{fig:hq}
\end{minipage}
\begin{minipage}[htbp]{0.8cm}
\end{minipage}
\begin{minipage}[htbp]{5.4cm}
\caption{$H_q$ for various Monte Carlo options}
 \label{fig:mc}
\end{minipage}
 \end{center} 
\vspace{-0.75cm}   
\end{figure}

The $H_q$ for the charged-particle multiplicity distribution of the full, 
light- and b-quark samples (figure \ref{fig:hq}) exhibit a first 
negative minimum at $q=5$ and quasi-oscillation for greater $q$. 
The $H_q$ measured from the light-quark sample are found to agree very 
well with those of the full sample, while slight differences exist,
mainly at low q, for the $H_q$ measured from the b-quark sample.
The observed behavior is similar to that predicted by the NNLLA.
However, also JETSET~\cite{jetset} agrees very well with all the data samples, even 
though the parton shower of JETSET does not use NNLLA. 
The same behaviour is found when using other parton generation and 
fragmentation models and even
when we use matrix element production of $\mathrm{q}\bar{\mathrm{q}}$ only,
even with independent fragmentation.
This shows us that the $H_q$ behavior can be reproduced without the need 
for the NNLLA of pQCD. 

Furthermore, our analysis of jet multiplicity obtained at perturbative energy scales 
($\gtrsim1\GeV$),
where pQCD predictions for the behaviour of $H_q$ should be directly accessible,
did not show any of the pQCD predictions made for the $H_q$~\cite{mystuff} .
Therefore, the $H_q$ behavior seen for 
the charged-particle and jet multiplicity distribution at
non-perturbative energy scales only, appears unrelated to the behavior of the $H_q$ 
calculated in NNLLA. 
\section{The phenomenological approach}
\begin{figure}
   \begin{center}
    \includegraphics[width=5.8cm,height=6cm]{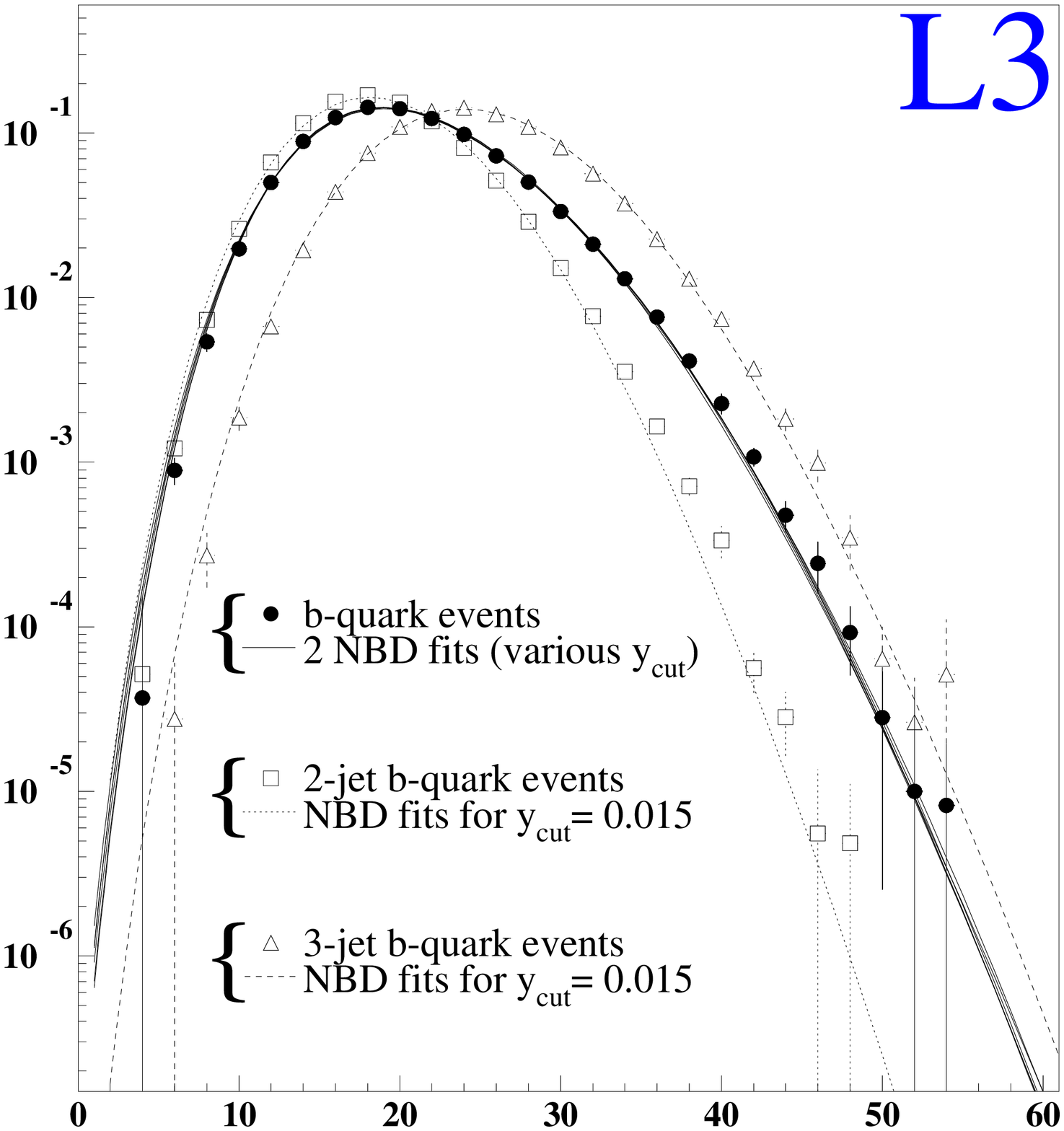}
    \includegraphics[width=5.8cm,height=6cm]{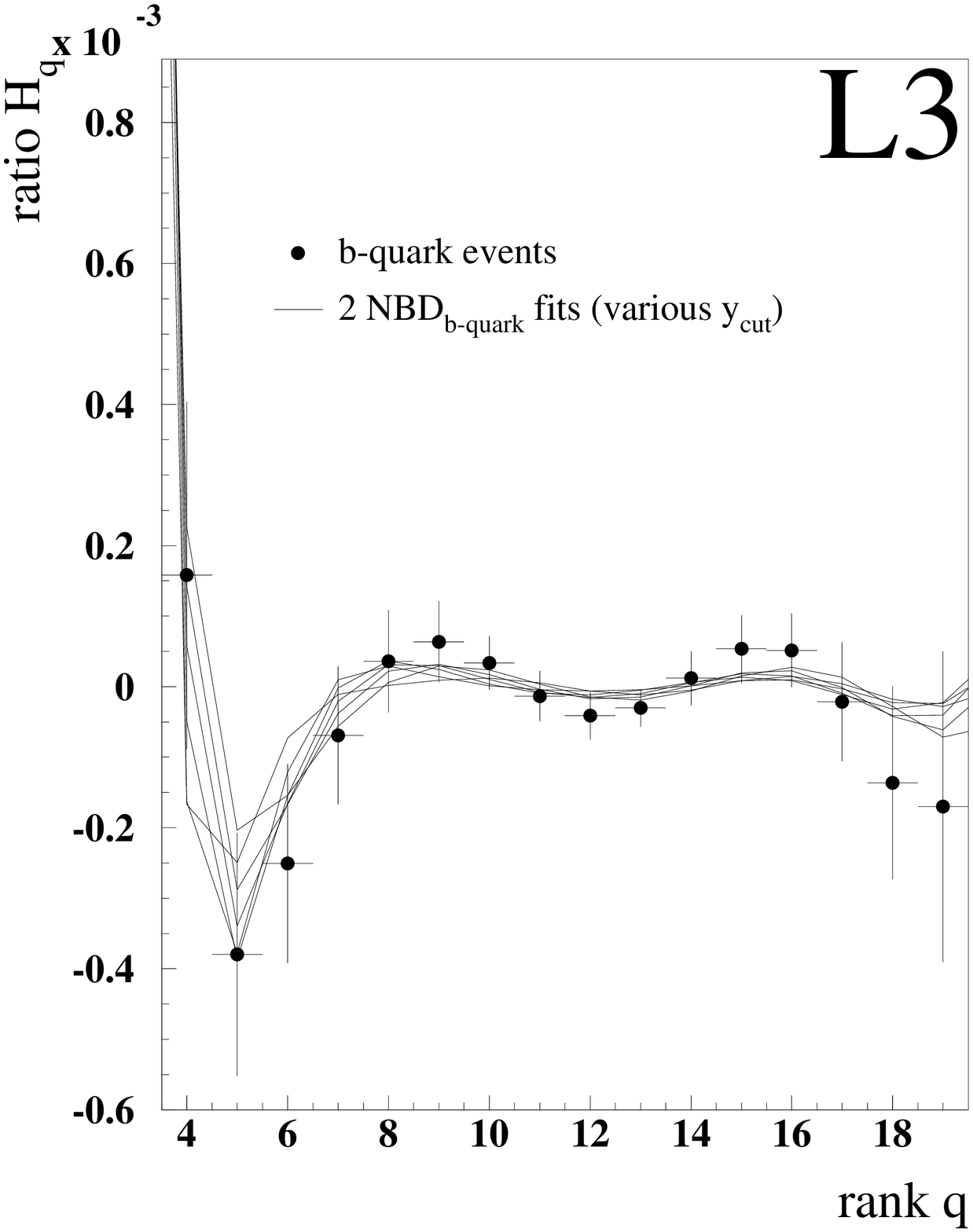}
\begin{minipage}[htbp]{5.4cm}
 \caption{Charged-particle multiplicity distributions for b-quark events and 2-jet and 
3-jet b-quark events, together with their parametrizations.}
 \label{fig:pnfit}
\end{minipage}
\begin{minipage}[htbp]{0.8cm}
\end{minipage}
\begin{minipage}[htbp]{5.4cm}
\caption{$H_q$ of the charged-particle multiplicity distribution for b-quark with 
the $H_q$ calculated from the 2NBD parametrizations}
 \label{fig:hqfit}
\end{minipage}
 \end{center}
\vspace{-0.75cm}
\end{figure}
This approach relies on the assumption that we can view the charged-particle 
multiplicity distribution as a superposition of distributions originating from 
various processes related to the topology of the event,
as 2-jet, 3-jet, light- or heavy-quark events.
Assuming that each of these processes can 
by itself be described by a relatively simple parametrization as the NBD, the charged-particle 
multiplicity distribution of the full sample would then be a weighted sum of all 
the contributions. All together, these various contributions would explain 
the shape of the charged-particle multiplicity distribution and, hence, the 
$H_q$ behavior.
We checked mainly two hypotheses with this parametrization.

1. The first assumes that the shape of the charged-particle multiplicity distribution 
of the full sample arises from the superposition of 2-jet and 3-jet events.
Our 2-jet and 3-jet samples were obtained using the Durham algorithm\cite{dur} for a 
set of six $y_\mathrm{cut}$ values.
As parameters for the NBD's, we used the means and dispersions calculated  
from the experimental 2-jet and 3-jet charged-particle multiplicity distributions.
The relative weight between the two NBD's was taken to be the 
fraction of 2-jet events for a given $y_\mathrm{cut}$ value. This gives us 
a fully constrained 2NBD parametrization of the full sample.
The resulting $\chi^2$ are given in the left half of table 1.
Since the $H_q$ moments from charged-particle multiplicity distributions 
of full, light- and b-quark samples are very similar, 
we also tested this hypothesis on light and b-quark samples separately, isolating in 
these cases the 2-jet and 3-jet events from the light- and b-quark samples.
The resulting $\chi^2$ for the b-quark sample are given in the right half of
table 1.
\begin{figure}
   \begin{center}
    \includegraphics[width=5.8cm,height=6cm]{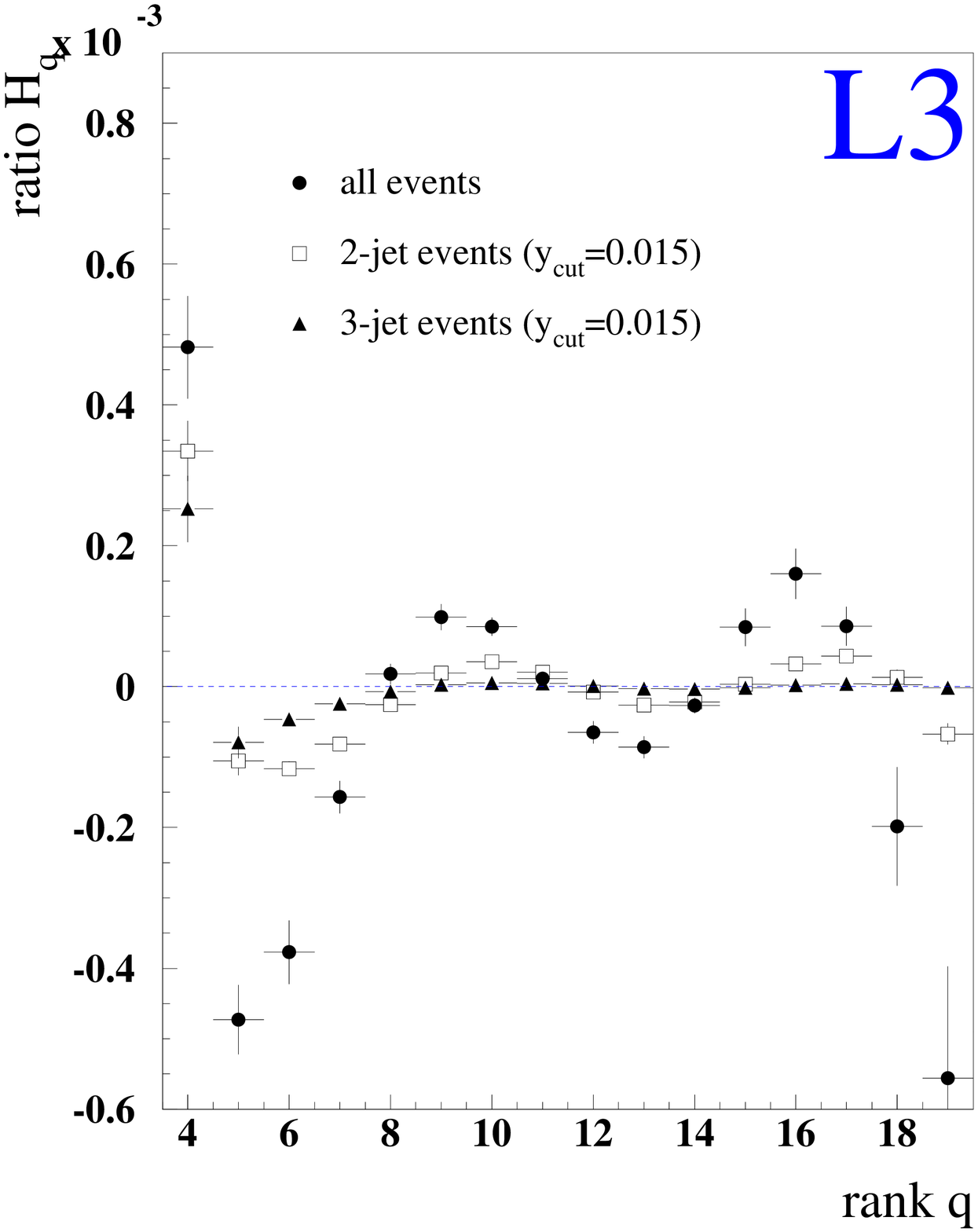}
    \includegraphics[width=5.8cm,height=6cm]{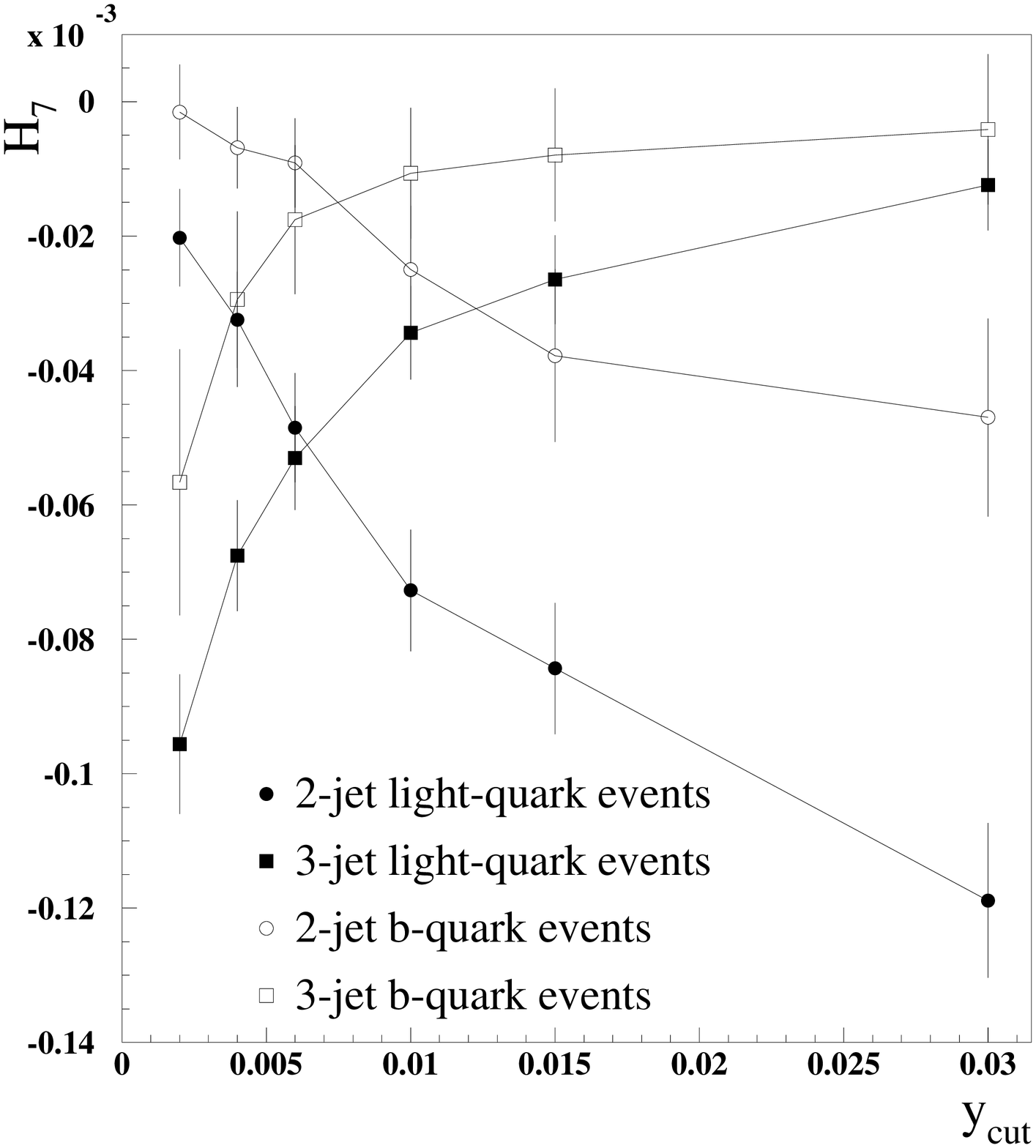}
\begin{minipage}[htbp]{5.4cm}
 \caption{$H_q$ of the charged-particle multiplicity distribution, for all events
and for 2-jet and 3-jet events ($y_\mathrm{cut}=0.015$)}
 \label{fig:hq23jet}
\end{minipage}
\begin{minipage}[htbp]{0.8cm}
\end{minipage}
\begin{minipage}[htbp]{5.4cm}
\caption{Evolution of $H_7$ as a function of $y_\mathrm{cut}$ for 2-jet and 3-jet, light- and
b-quark events}
 \label{fig:h7}
\end{minipage}
 \end{center}
\vspace{-0.75cm}
\end{figure}
We find amazingly good $\chi^2$ for the 2NBD parametrizations of the  full 
(column 2 of table 1), light- (not shown) and b-quark (column 5 of table 1) samples.
(see also figure~\ref{fig:pnfit} for the case of the b-quark sample).
We calculated the $H_q$ from 
the parametrizations and also these are found to be in good agreement with the $H_q$
measured for the full, light and b-quark samples (figure~\ref{fig:hqfit}
for the b-quark sample).
However, none of the NBD parametrizations are able to describe any of the 
individual 2- or 3-jet 
charged-particle multiplicity distributions themselves, 
even though the $\chi^2$ is seen to decrease when the purity increases.
2. We also attempted to parametrize the 2-jet and 3-jet charged-particle 
multiplicity distributions, and as a consistency check the full sample, 
by a superposition of light and b-quark events, using in that case as relative
weight between the two NBD's, the fraction of b-quark events, $R_b$\cite{lephf}.
Results are summarized in table 2. 
We don't find any agreement at all, neither for 2NBD parametrization of 
the full sample which has a $\chi^2/\mathrm{dof}$ near $13$.
This 
constitutes by its failure a good check of the method, since it shows that not 
all combinations of two NBD's agree with the data.
We extended the study of the 2-jet and 3-jet samples to the measurement of their 
$H_q$ moments. We find that, even if the oscillations are still there, 
their amplitudes are far smaller than for the full sample (figure~\ref{fig:hq23jet}).
Furthermore the size of the oscillations decreases when 
the purity in 2 jet (3 jet) in the 2-jet (3-jet) sample increases.
Differences at low $q$ (mainly for $q<8$) are found between the $H_q$ of 2-jet 
and 3-jet events, and also large differences are seen for fixed $q$ when the $H_7$ 
moment is plotted versus $y_\mathrm{cut}$ (figure~\ref{fig:h7}). 
We see differences between $H_q$ of 2-jet and 3-jet events of the 
light- and b-quark samples, however the oscillations are comparable to those of  
2-jet or 3-jet events of the full samples.
All together, this supports the phenomenological approach when we assume that
the shape of the charged-particle multiplicity distribution arises from
a superposition of 2-jet and 3-jet events.
\begin{table}
\caption[]{$\chi^2$ between the 2-jet, 3-jet parametrization and their 
experimental counterpart}
\begin{center}
\begin{tabular}{|c|c|c|c||c|c|c|}\cline{2-7}
  \multicolumn{1}{c|}{} &
 \multicolumn{3}{c||}{$\chi^2/\mathrm{dof}$ for the full sample} &
 \multicolumn{3}{c|}{$\chi^2/\mathrm{dof}$ for the b-quark sample}\\
\hline 
$y_\mathrm{cut}$& 
$\mathrm{2NBD}_\mathrm{all}$&
$\mathrm{NBD}_\mathrm{2jet}$&
$\mathrm{NBD}_\mathrm{3jet}$&
$\mathrm{2NBD}_\mathrm{all}$&
$\mathrm{NBD}_\mathrm{2jet}$&
$\mathrm{NBD}_\mathrm{3jet}$\\
\hline
$0.03$     & 1.4  & 56  & 8.5 & 2.6  & 19  & 2.2 \\
\hline
$0.015$    & 0.6  & 36  & 15  & 1.4  & 13  & 3.4 \\
\hline
$0.01$     & 20.4 & 8   & 6.3 & 0.93 & 9.8 & 4.4 \\
\hline
$0.006$    & 0.7  & 16  & 36  & 0.53 & 5.5 & 6.4 \\
\hline
$0.004$    & 1.5  & 9.4 & 57  & 0.53 & 3.5 & 8.5 \\
\hline
$0.002$    & 5.   & 5.5 & 119 & 3    & 0.8 & 15  \\
\hline
\end{tabular}\end{center}
\label{tab:23jet}
\vspace{-0.5cm}
\end{table}
\begin{table}
\caption[]{$\chi^2$ between the light-, b-quark parametrization and their 
experimental counterpart}
\begin{center}
\begin{tabular}{|c|c||c|}\cline{2-3}
  \multicolumn{1}{c|}{} &
  \multicolumn{2}{c|}{$\chi^2/\mathrm{dof}$} \\
\hline 
$y_\mathrm{cut}$& 
$\mathrm{2NBD}_\mathrm{2jet}$&
$\mathrm{2NBD}_\mathrm{3jet}$\\
\hline
$0.03$     & 74  & 8   \\ 
\hline
$0.015$    & 46  & 14  \\
\hline
$0.01$     & 39  & 9   \\ 
\hline
   $0.006$    & 23  & 33  \\
\hline
$0.004$    & 14  & 48  \\
\hline
$0.002$    & 5   & 100 \\
\hline
\end{tabular}\end{center}
\label{tab:lbtag}
\vspace{-0.5cm}
\end{table}
\section{Conclusions}
The oscillatory behavior of the $H_q$ moments of the charged-particle multiplicity
distribution is usually interpreted as a confirmation of NNLLA, but investigations 
performed on different models of parton generation and 
for different fragmentation models have shown similar oscillatory behavior in all 
cases. Furthermore, the analysis of the $H_q$ of the jet multiplicity 
distributions reveals that this behavior appears only for very small 
$y_{\mathrm{cut}}$, corresponding to energy scales $\lesssim 100\MeV$, 
far from the perturbative region. This gives us strong indications that the 
oscillatory behavior is not related to the behavior predicted by the NNLLA.
In search of an alternative origin of this $H_q$ behavior we have, therefore, 
investigated a more phenomenological answer which assumes that the shape of the 
multiplicity distribution results from the superposition of 2-jet and 3-jet events. 
Using a weighted sum of 2 NBD's as a parametrization,
we found very good agreement for both the charged-particle multiplicity 
distribution and its $H_q$ moment. This supports the idea that the 
main feature in the shape of the charged-particle multiplicity still 
visible in the final states are due to the presence of hard gluon radiation
and to the hadronization.

\end{document}